\documentclass[aps,prb,twocolumn,groupedaddress]{revtex4-1}
\usepackage{amsmath,amssymb}
\usepackage{graphicx}
\usepackage{bm}
\usepackage{epstopdf} 

\AppendGraphicsExtensions{.tif}

\newcommand{\BS}{Bi$_2$Se$_3$}
\newcommand{\BT}{Bi$_2$Te$_3$}

\newcommand{\BSTS}{(Bi$_{1-x}$Sb$_x$)$_2$(Te$_{1-y}$Se$_y$)$_3$}

\newcommand{\MBT}{Mn-\BT}

\newcommand{\TC}{$T_{\rm{C}}$}

\renewcommand{\vec}[1]{\mbox{\boldmath$1$}}

\newcounter{lastnote}

\def\bc{\begin{center}}
\def\ec{\end{center}}
\def\be{\begin{equation}}
\def\ee{\end{equation}}
\renewcommand{\vec}[1]{\mbox{\boldmath$1$}}

\begin{document}
\title{Ferromagnetism and Spin-dependent Transport in n-type Mn-\BT~Thin Films}
\author{Joon Sue Lee$^1$, Anthony Richardella$^1$, David W. Rench$^1$, Robert D. Fraleigh$^1$, Thomas C. Flanagan$^1$, Julie A. Borchers$^2$, Jing Tao$^3$ and Nitin Samarth$^1$}
\email{nsamarth@psu.edu}
\affiliation{$^1$Department of Physics, The Pennsylvania State University, University Park, Pennsylvania 16802, USA}
\affiliation{$^2$National Institute of Standards and Technology, Gaithersburg, Maryland 20899, USA}
\affiliation{$^3$Brookhaven National Laboratory, Upton, New York 11973 USA}

\date{\today}
\begin{abstract}
We describe a detailed study of the structural, magnetic, and magneto-transport properties of single-crystal, n-type, Mn-doped \BT~thin films grown by molecular beam epitaxy. With increasing Mn concentration, the crystal structure changes from the tetradymite structure of the \BT~parent crystal at low Mn concentrations towards a BiTe phase in the (\BT)$_m$(Bi$_2 )_n$ homologous series. Magnetization measurements reveal the onset of ferromagnetism with a Curie temperature in the range 13.8 K $-$ 17 K in films with $\sim$ 2 $-$ $\sim$ 10 \% Mn concentration. Magnetization hysteresis loops reveal that the magnetic easy axis is along the c-axis of the crystal (perpendicular to the plane). Polarized neutron reflectivity measurements of a 68 nm-thick sample show that the magnetization is uniform through the film. The presence of ferromagnetism is also manifest in a strong anomalous Hall effect and a hysteretic magnetoresistance arising from domain wall scattering. Ordinary Hall effect measurements show that the carrier density is n-type, increases with Mn doping, and is high enough ($\geq 2.8 \times 10^{13}$ cm$^{-2}$) to place the chemical potential in the conduction band. Thus, the observed ferromagnetism is likely associated with both bulk and surface states. Surprisingly, the Curie temperature does not show any clear dependence on the carrier density but does increase with Mn concentration. Our results suggest that the ferromagnetism probed in these Mn-doped \BT~films is not mediated by carriers in the conduction band or in an impurity band.
\end{abstract}
\maketitle

\section{Introduction}
Three-dimensional (3D) topological insulators (TIs) have attracted much interest due to the presence of spin-momentum-locked surface states that are protected by time reversal symmetry (TRS).\cite{Moore2010,Hasan2010,Qi2010,Fu2007,Hsieh2008} Magnetic doping of a TI breaks TRS and is predicted to open a gap at the Dirac point,\cite{Qi2008,Liu2009,Chen2010,Wray2011} resulting in exotic quantum phenomena such as the topological magneto-electric effect,\cite{Qi2008} the induction of a magnetic monopole,\cite{Qi2009} and the quantized anomalous Hall effect.\cite{Yu2010,Chang2013QAHE} Recent experimental progress towards incorporating ferromagnetism with 3D TIs has indeed led to the realization of such effects, including the observation of a hedgehog spin texture,\cite{Xu2012} gate-tunable ferromagnetism,\cite{Checkelsky2012} and a quantized anomalous Hall effect.\cite{Chang2013QAHE} This interplay between magnetism and topologically protected states is of potential interest for applications in quantum computing and spintronics.\cite{Pesin2012} 

Given this context, there is a strong motivation for exploring the magnetic behavior of transition metal doped materials such as Bi-chalcogenides which have been established as hosting TI surface states. Indeed, ferromagnetism was studied in a variety of transition metal doped chalcogenide compounds well before the interest in these materials as 3D TIs. For instance, ferromagnetic Curie temperatures (\TC) ranging up to 190 K were reported in bulk crystals of \BT~doped with V, Cr, Mn,\cite{Dyck2002,Dyck2005,Choi2005} and Fe,\cite{Kulbachinskii2002} as well as in thin films grown by molecular beam epitaxy (MBE).\cite{Chien2007} More recent studies of magnetically-doped chalcogenides, primarily driven by an interest in topological phenomena, have shown ferromagnetism in a variety of samples: bulk crystals of p-type, Mn-doped \BT~\cite{Hor2010} and Fe-doped \BS,\cite{Choi2011} as well as thin films of Cr-doped Bi$_2$Te$_3$,\cite{Bao2013} Cr-doped \BS,\cite{Haazen2012} n-type, Mn-doped \BS,\cite{Zhang2012} and Cr-doped \BSTS.\cite{Zhang2013,Chang2013} Other magnetic phases (paramagnetic, antiferromagnetic, spin glass) have also been observed.\cite{Choi2005,Choi2011} Before drawing detailed conclusions about the influence of ferromagnetism on surface states, careful structural and magnetic characterization of these types of samples is needed. For instance, we recently showed that nanoscale segregation of magnetic atoms can occur at the surface of thin films of Mn-doped \BS,\cite{Zhang2012}although macroscopic probes of ferromagnetism such as ferromagnetic resonance and polarized neutron reflectivity are consistent with a uniform bulk magnetization.\cite{VonBardeleben2013}

In this study, we demonstrate robust bulk ferromagnetism in n-type, Mn-doped \BT~thin films grown by MBE. We note that at low Mn-doping, these films are n-type and in the tetradymite phase but still show similar values of \TC~as the p-type, tetradymite bulk crystals of Mn-\BT~studied in the past.\cite{Hor2010} At higher Mn-doping, the crystalline phase changes to one in which some \BT~quintuple layers are separated by Bi bilayers, similar to earlier reports of ferromagnetic, n-type bulk crystals.\cite{Bos2006} We start by describing the MBE synthesis of the Mn-doped \BT~thin films in section II. The following section discusses the crystal structure of the MBE-grown films, characterized by x-ray diffraction (XRD) and high-resolution transmission electron microscopy (HRTEM). We then discuss magnetization measurements using superconducting quantum interference device (SQUID) magnetometry and polarized neutron reflectivity (PNR) in section IV. Section V discusses measurements of magneto-transport and magnetic anisotropy that are consistent with the results of SQUID magnetometry. Finally, we compare the films to bulk samples and discuss the nature and origin of ferromagnetism in the Mn-doped \BT~films in section VI. 

\begin{figure*}
\includegraphics[width=180mm]{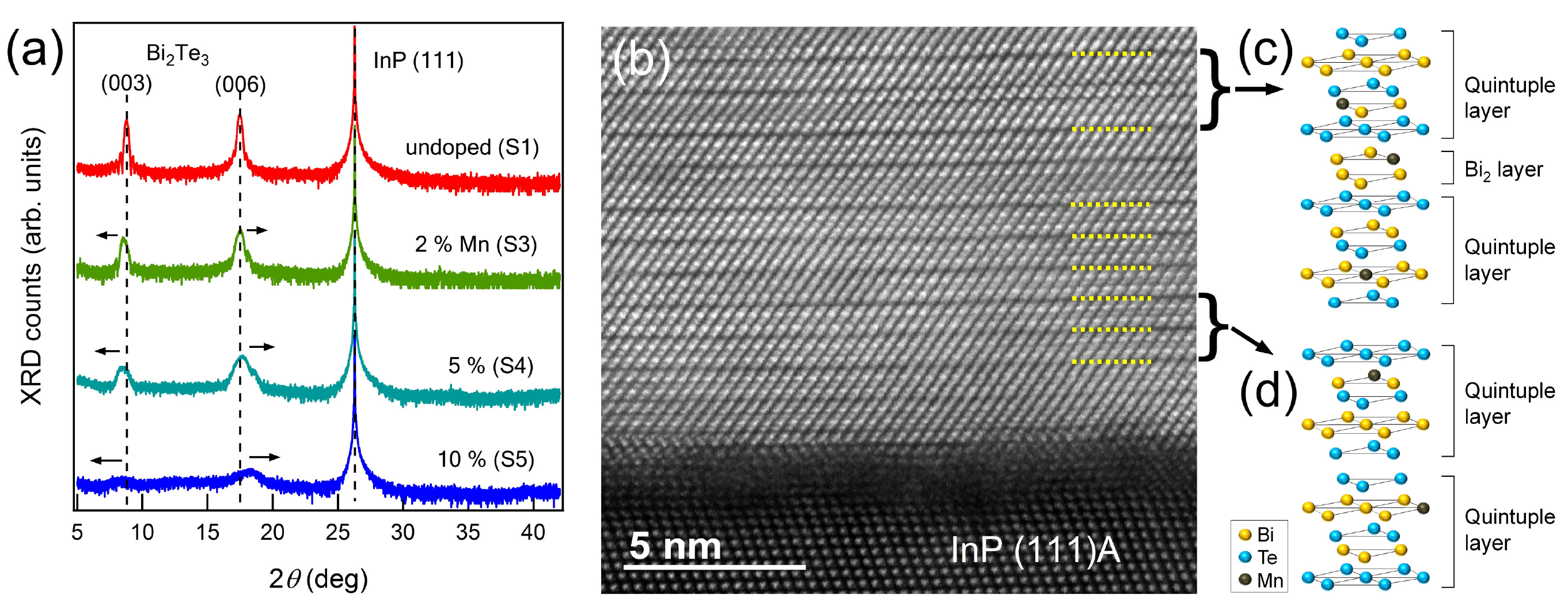} 
\caption{(Color online) (a) XRD of undoped \BT~(S1) and Mn-doped \BT~thin films (S3, S4, and S5). (003) and (006) peaks of \BT~shift in opposite directions (black arrows) with increasing Mn concentration while the (111) peak of InP does not change with Mn-doping. Curves are offset for clarity. (b) A HAADF-STEM image of a Mn-doped \BT~thin film on an InP (111)A substrate with 5 \% Mn concentration (S4). Dotted yellow lines indicate QLs and unit layers composed of a Bi bilayer sandwiched between two QLs. The atomic crystal structures of (c) a QL-Bi$_{2}$-QL unit layer and (d) two \BT~QLs, with Bi partially substituted by Mn.}
\label{structure}
\end{figure*}

\begin{table*}[t]
	\caption{List of undoped and Mn-doped \BT~thin films with different Mn concentrations. Thickness, beam equivalent pressure ratio (Bi/ Mn), Mn atomic percentage (20 \% relative error) measured and calculated by RBS and SIMS, 2D carrier concentration ($n^{2D}$), mobility ($\mu$), existence of AHE by magneto-transport measurements, onset temperatures of AHE ($\pm 1$ K error), and \TC~by SQUID magnetometry for each sample are shown. }
  \begin{tabular*}{1\textwidth}{@{\extracolsep{\fill} } c c c c c c c c c }
  \hline \hline
  Sample & t (nm) & Mn/Bi & Mn (\%) & $n^{2D}$ $(10^{13} cm^{-2})$ & $\mu$ $(cm^2/Vs)$ & AHE & \TC~(AHE) & \TC~(SQUID) \\ \hline
  S1 & 28 & 0 & 0 & 2.90 & 620 & X & - & - \\ 
  S2 & 30 & 0.02 & 1 & 3.21 & 196 & X & - & -  \\ 
  S3 & 30 & 0.04 & 2 & 10.1 & 98.2 & O & 11 K & 13.8 K \\ 
  S4 & 30 & 0.07 & 5 & 14.5 & 79.2 & O & 15 K & 15.0 K \\ 
  S5 & 30 & 0.13 & 10 & 6.51 & 73.0 & O & 16 K & 17.0 K \\ 
  S6 & 68 & 0.04 & 4.5 & 57.1 & 107 & O & 10 K & 15.0 K \\ \hline \hline
  \end{tabular*}
\label{samples}
\end{table*}

\section{Sample Synthesis}
The Mn-doped \BT~thin films were epitaxially grown on InP (111)A substrates by MBE using high purity elemental Bi, Mn, and Te in ultra-high vacuum ($\sim$ $10^{-10}$ Torr). Despite the van der Waals bonding to the substrate, the TI film quality is generally improved by reducing the lattice mismatch between the film and the substrate. InP(111)A was chosen because it has an in-plane lattice constant (4.150 \AA) which is closer to \BT~(4.380 \AA) than GaAs(111) (3.998 \AA), Si(111) (3.840 \AA), or sapphire (4.751 \AA). The InP oxide was desorbed under a Se flux prior to growth resulting in an approximately 1 nm thick amorphous region at the interface, as seen in Fig.\ref{structure}. To explore the role of the Mn concentration on the magnetic properties of Mn-Bi$_2$Te$_3$, thin films of a wide range of Bi to Mn beam equivalent pressure ratios were synthesized while maintaining a film thickness of $\sim 28 - 30$ nm. In addition, we grew a 68 nm sample for PNR measurements where the InP was desorbed under an As flux. The set of samples and their properties are listed in Table \ref{samples}. 

\section{Crystal Structure}

The atomic compositions of Mn-doped \BT~films S3, S5, and S6 were analyzed by a secondary ion mass spectrometry (SIMS) depth profiles. The total Mn concentration in the highly doped film (S5) was determined by Rutherford backscattering (RBS), to be 10 atomic percent with 20 \% relative error. This was then used to calibrate the SIMS measurements. Samples S1 through S5 were similar thicknesses and grown under similar conditions. The Mn concentration of S3 was found to be $\sim$2 \% from SIMS and x-ray photoelectron spectroscopy (XPS). The Mn concentrations of films S1 and S4 were then estimated to be 1 \% and 5 \%, respectively, using the Mn/Bi beam-equivalent-pressure ratios relative to S3 and S5. Sample S6 was thicker and was grown with a slightly different geometry in the chamber, resulting in a higher Mn concentration (4.5 \%) than would have been estimated just from the Mn/Bi pressure ratio relative to the other samples. 

The crystal structure of undoped and Mn-doped \BT~films was studied by XRD and TEM. In Fig. 1(a), the XRD pattern of the undoped film shows sharp \BT~(003) and (006) peaks. Mn-doping broadens both peaks and, interestingly, causes the (003) and (006) peaks to shift in opposite directions, with shoulders forming at high Mn concentrations, indicating changes in the crystal structure. The crystal structure of the Mn-doped \BT~film with 5 \% Mn (S4) was directly imaged by TEM. A basic structural unit of rhombohedral \BT~(tetradymite phase) is a quintuple layer (QL) which consists of five alternating atomic layers, Te-Bi-Te-Bi-Te. Heavy elements, such as Bi, scatter the electron beam to large angles more effectively than lighter elements, such as Te, allowing the high angle annular dark field (HAADF) scanning transmission electron microscopy (STEM) image in Fig.\ref{structure}(b) to show this elemental arrangement within the repeated QL structure as dim-bright-dim-bright-dim (corresponding to Te-Bi-Te-Bi-Te, as expected). However, a modified structure with additional atomic layers is seen in the upper part of the image. Consistent with the atomic contrast, this most likely corresponds to the formation of a Bi bilayer between two QLs, as reported in earlier studies of single crystals of bulk, Mn-doped BiTe.\cite{Bos2006} This crystalline phase is distinct from tetradymite \BT~and is a member of the (\BT)$_m$(Bi$_2 )_n$ homologous series. The shifts in the XRD are also consistent with such Bi-rich phases being present. Figures\ref{structure}(c) and (d) show the atomic crystal structure of a QL-Bi$_{2}$-QL unit layer and two QLs, respectively. Larger scale HAADF-STEM images suggest that such Bi bilayers are scattered randomly in different regions of the film. Similar to the case of Bi bilayers in \BS, we expect such a structure to retain its topological character.\cite{Valla2012} Mn likely substitutes for Bi in both locations, but from electron energy loss spectroscopy (EELS) line scans an interstitial position cannot be ruled out. From these TEM and XRD studies, we deduce that adding Mn into \BT~results in the formation of Bi bilayers, with the crystal structure gradually transitioning from pure tetradymite to (\BT)$_m$(Bi$_2 )_n$ with $n/m$ approaching 0.5 at high Mn concentrations. As we show later, this likely also leads to n-doping. 

\begin{figure}
\includegraphics[width=80mm]{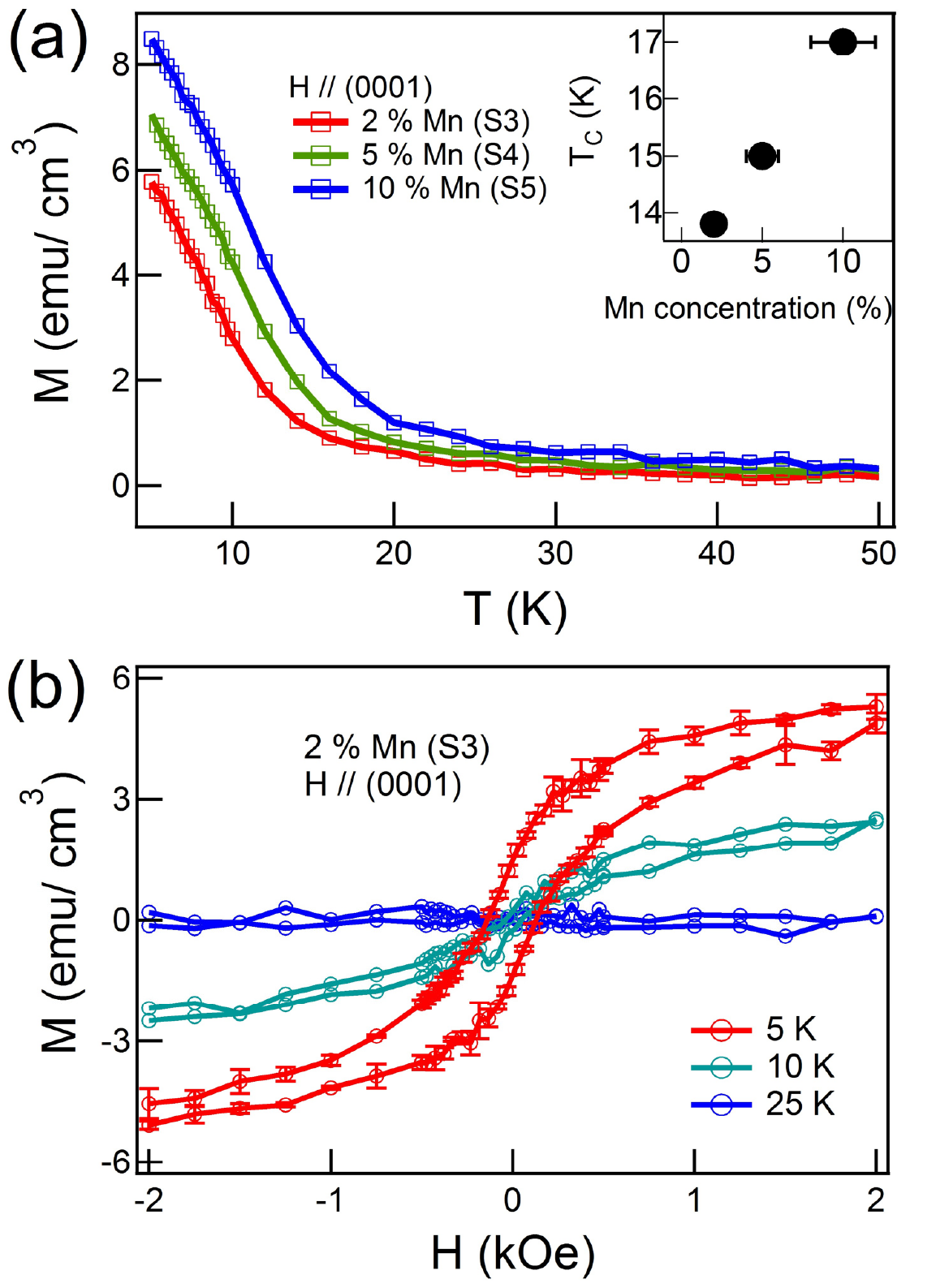} 
\caption{(Color online) SQUID magnetometry. (a) Temperature dependence of magnetization with perpendicular-to-the-plane magnetic field for Mn-doped \BT~thin films with 2 \% (S3), 5 \% (S4), and 10 \% (S5) Mn concentration. Inset plots \TC~as a function of Mn concentration. (b) Magnetic field sweep of magnetization with perpendicular-to-the-plane magnetic field for a Mn-doped \BT~thin film with 2 \% Mn concentration (S3). Note that the background magnetization of the InP substrate has been subtracted in (b). }
\label{SQUID}
\end{figure}

\begin{figure*}
\includegraphics[width=180mm]{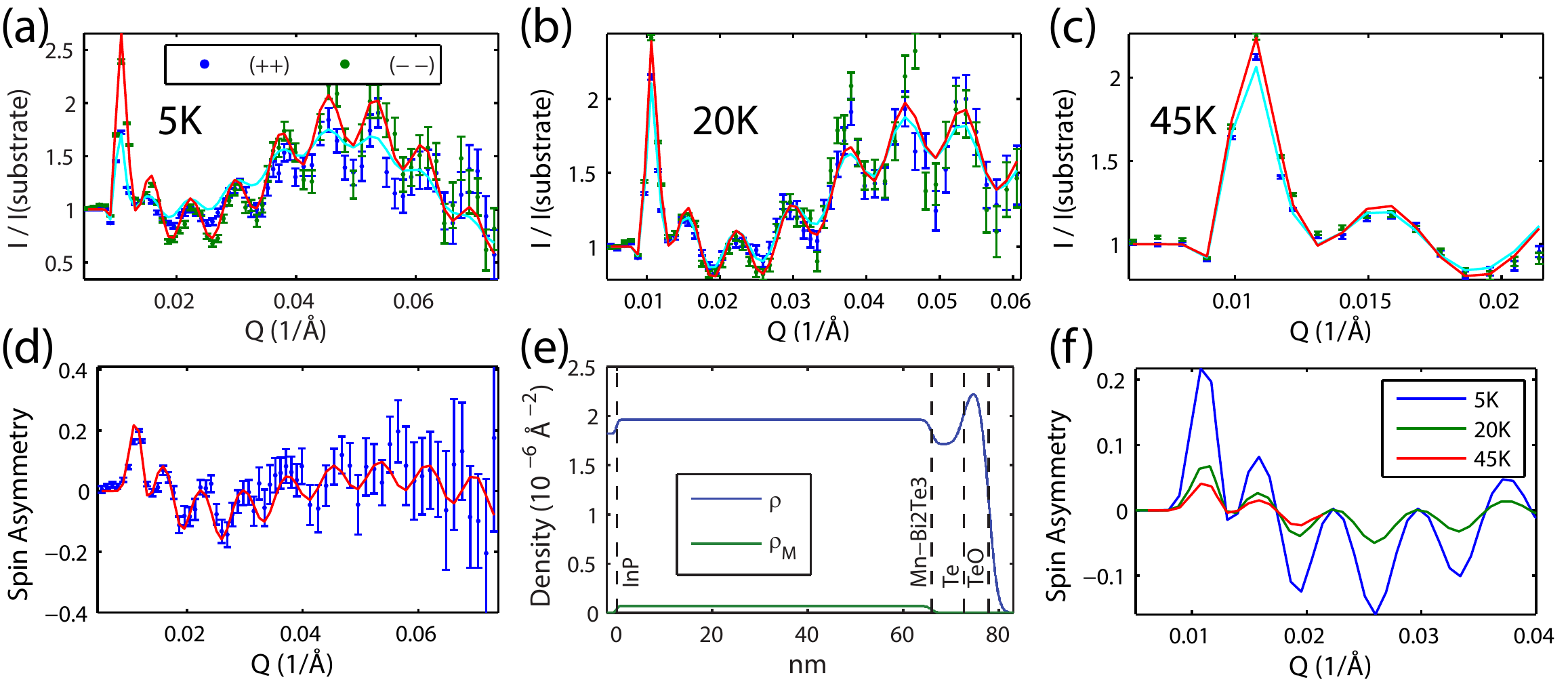} 
\caption{(Color online) PNR magnetometry measured with an in-plane 8 kOe magnetic field. Reflectivity curves and fits (solid lines) for up  ($++$) and down  ($--$) spin neutrons are plotted on a normalized Fresnel scale at (a) 5 K, (b) 20 K and (c) 45 K. (d) Spin asymmetry and fit at 5 K. (e) Structural and magnetic scattering length densities from the 5 K fits for each layer of the sample. The magnetization, proportional to $\rho_M$ (green curve), is uniform across the \MBT~layer. (f) Fits to spin asymmetry show a small but measurable magnetization persisting to 45 K, well above the \TC~measured by SQUID magnetometry.}
\label{PNR}
\end{figure*}

\section{Magnetic Characterization}
SQUID magnetometry reveals the onset of ferromagnetism in the Mn-doped \BT~thin films for Mn concentrations larger than 1 \%. Figure\ref{SQUID}(a) shows the temperature dependence of the remanent magnetization in films with 2 \% (S3), 5 \% (S4) and 10 \% (S5) Mn concentrations, measured above 5 K after field-cooling in a 9 kOe magnetic field perpendicular to the plane (H // (0001)). The \TC~of each film was determined by finding the steepest slope and extrapolating a linear fit to zero magnetization. Table \ref{samples} and the inset to Fig.\ref{SQUID}(a) show that the \TC~of each is 13.8 K, 15.0 K and 17.0 K, respectively. Focusing on S2 (Fig.\ref{SQUID}(b)), below \TC~we observe ferromagnetic hysteresis of the magnetization with a perpendicular magnetic field. No significant hysteresis was observed with an in-plane magnetic field (H // (1100)). Similar hysteretic behavior was observed in the other samples, confirming that the magnetic easy axis is out-of-plane, along the c-axis of the Mn-\BT~crystal. 

Sample S6 was further studied by PNR at the NIST Center for Neutron Research (NCNR) on the Polarized Beam Reflectometer (PBR). This sample was thicker than the other samples in this study ($\sim 68$ nm) and was capped with Te to protect the surface from exposure to atmosphere. The sample was field cooled with an in-plane applied field of 8 kOe. The in-plane external field is necessary because, due to the neutron selection rules, the sample magnetization has to be aligned in-plane to observe the magnetic structure. Measurements were then made at 5 K, 20 K, and 45 K (Fig.\ref{PNR}) with a neutron wavelength of $\lambda$ = 4.75 \AA. Incident neutrons were polarized parallel ($+$) or anti-parallel ($-$) to the applied field and the reflected ($++$), ($--$) and spin flip ($+-$), ($-+$) intensities were measured. These reflectivity curves include scattering contributions from both the structure and magnetism of the sample and were modeled to extract the parameters that best fit them using REFL1D.\cite{Kienzle2011} The spin flip intensities were insignificant, indicating that the sample magnetization was aligned with the external field. The difference between the ($++$) and ($--$) reflectivities divided by their sum is called the spin asymmetry and is related to the magnetization of the sample as a function of depth. The sample was further measured by x-ray reflectivity (XRR) to help determine the layer thicknesses used in modeling the fits to the PNR data. The Mn-doped \BT~layer was found to have a uniform scattering length density and magnetization through the thickness of the film, Fig.\ref{PNR}(e) with a magnetization of 23.2 emu/cm$^3$, within about a factor of two of that measured from SQUID for this sample. No enhanced magnetism or dead layers at the TI surfaces were needed to model the data. Surprisingly, a highly reduced but measurable uniform magnetization was observed up to 45 K, well above the \TC~measured by SQUID and transport measurements. We speculate that the PNR measurement may be averaging over small, randomly distributed regions in the sample that have a higher \TC~than the bulk of the film and which were being aligned by the external field.

\begin{figure*}
\includegraphics[width=180mm]{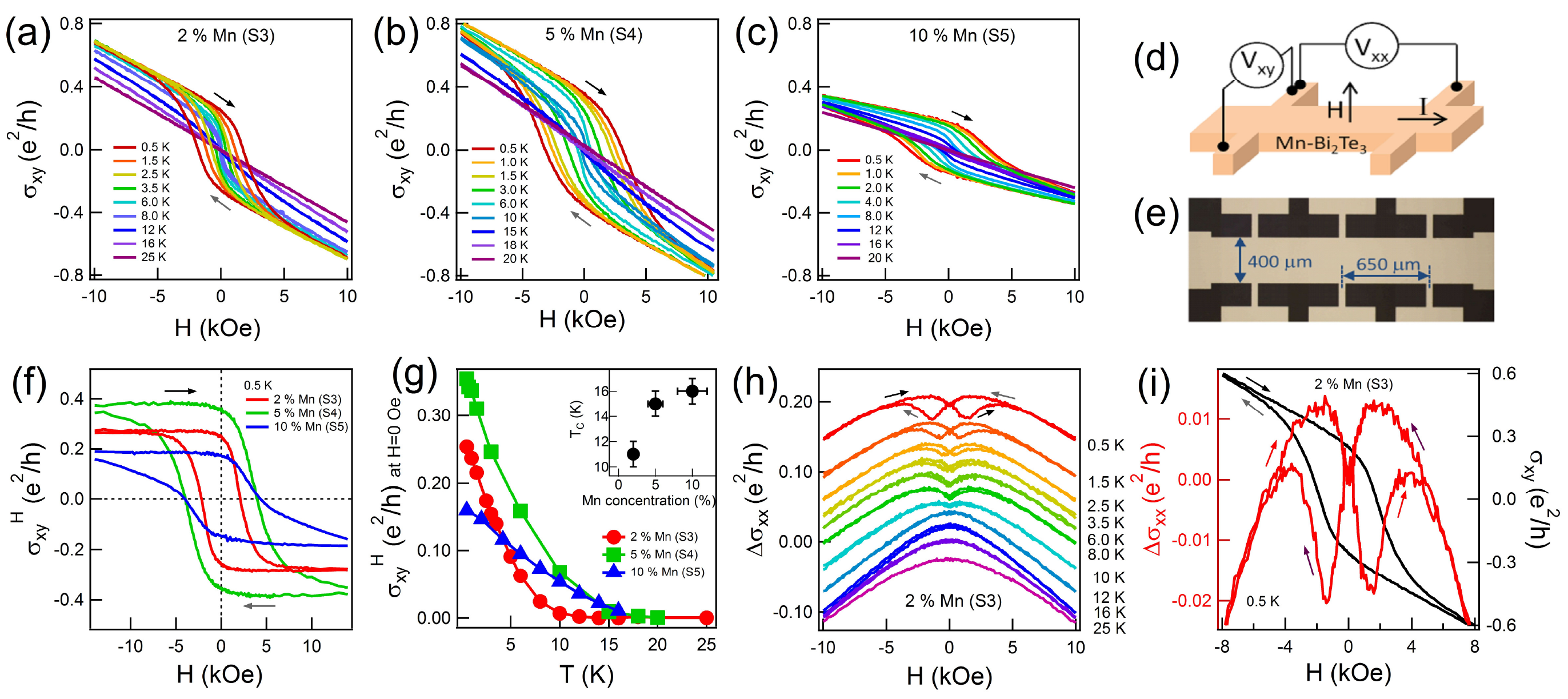} 
\caption{(Color online) Hall and longitudinal conductivity obtained from magneto-transport measurements. A schematic of the measurement setup and a photo image of a Hall bar are shown in (d) and (e), respectively. Panels (a)-(c) shows the temperature dependence of the Hall conductivity $\sigma_{xy}$ for Mn-doped \BT~films with 2 \% Mn (S3), 5 \% Mn (S4), and 10 \% (S5) Mn concentration. (f) Anomalous Hall conductivity $\sigma_{xy}^H$ at 0.5 K for various Mn concentrations (S3, S4, and S5). (g) Temperature dependence of $\sigma_{xy}$ at zero magnetic field with different Mn concentrations (S3: red circles, S4: green squares, and S5: blue triangles). The inset plots \TC~as a function of the Mn concentration. Black solid line is a linear fit. (h) Temperature dependence of longitudinal MC $\Delta\sigma_{xx}$ from sample S3. Ferromagnetic hysteresis is shown below 4 K. (i) Longitudinal conductivity $\Delta\sigma_{xx}$ (red) and Hall conductivity $\sigma_{xy}$ (black) at 0.5 K.}
\label{MR}
\end{figure*}

\section{Magneto-conductivity and Hall Effect Measurements}

The carrier density and mobility of the thin films were determined by magneto-transport measurements using Hall bars with 650 $\mu$m $\times$ 400 $\mu$m channels patterned by conventional photolithography and dry-etched in an Ar/Cl plasma. Standard four-probe lock-in measurements at 19 Hz were conducted with perpendicular-to-plane magnetic fields up to 6 T at temperatures down to 0.5 K. A schematic of the measurement setup and an optical microscope image of a typical device are shown in Figs. 3(d) and (e), respectively. The undoped \BT~film (S1) is n-type with a 2D carrier density of $2.9\times10^{13}$ cm$^{-2}$ at 4.2 K. The carrier density with the Fermi energy $E_{F}$ located at the bottom of the conduction band is estimated to be $2.8\times10^{13}$ cm$^{-2}$, calculated from the Fermi surface at the conduction band bottom of \BT~obtained by ARPES.\cite{Chen2009} Thus, $E_{F}$ of the undoped \BT~(S1) is located just above the bottom of the conduction band. With increasing Mn-doping, the carrier density increases and the chemical potential ($E_{F}$) of the Mn-doped films is in the conduction band (Table 1). The n-type conductivity and the position of $E_{F}$ in undoped and Mn-doped \BT~thin films were confirmed by the infrared (IR) optical measurements of samples of similar composition (but with thickness close to that of S6).\cite{Chapler2013} The carrier concentration of sample S6 is somewhat larger than the similarly doped S3 due to its conductive Te capping layer. 

We observed a strong anomalous Hall effect (AHE) in the samples with Mn concentrations of 2 \% or higher (S3, S4, and S5). Hall conductivity $\sigma_{xy}$ versus perpendicular magnetic field curves at various temperatures reveal the evolution of the AHE. Figures\ref{MR}(a) -- (c) show the temperature dependence of $\sigma_{xy}$ for S3, S4, and S5, respectively. At temperatures much higher than \TC, $\sigma_{xy}$ shows a linear dependence on field, but as the temperature decreases the curve becomes non-linear and then hysteretic below \TC. The onset temperature of the AHE is at around 11 K for 2 \% Mn concentration (S3). The coercive field gradually increases as the temperature decreases, reachinig about 1.8 kOe at 0.5 K in sample S2. The Hall conductivity $\sigma_{xy}$ is calculated from longitudinal resistivity $\rho_{xx}$ and Hall resistivity $\rho_{xy}$ as $\sigma_{xy}=\frac{\rho_{xy}}{(\rho_{xy}^2+\rho_{xx}^2)}$, and $\sigma_{xy}$ is composed of the normal Hall conductivity $\sigma_{xy}^N$ and the anomalous Hall conductivity $\sigma_{xy}^H$ as $\sigma_{xy}=\sigma_{xy}^N+\sigma_{xy}^H$. By subtracting the linear ordinary Hall conductivity $\sigma_{xy}^N$ from $\sigma_{xy}$, one can obtain the anomalous Hall conductivity $\sigma_{xy}^H$. In Fig.\ref{MR}(f), we plot the anomalous Hall conductivity $\sigma_{xy}^H$ versus perpendicular magnetic field at 0.5 K for samples with different Mn concentrations (S3, S4, and S5). The coercive field is larger with higher Mn concentration, but the saturated $\sigma_{xy}^H$ value is the largest with the intermediate doping (S4). In Fig.\ref{MR}(g), the temperature dependence of $\sigma_{xy}^H$ at zero magnetic field from S3, S4, and S5 shows that the onset temperature of the AHE hysteresis increases with Mn-doping and is consistent with the result from the SQUID measurements.

Besides the AHE, the ferromagnetism in the Mn-doped \BT~can also be seen from the longitudinal magneto-conductivity (MC) with a magnetic field perpendicular to the plane. For the film with 2 \% Mn concentration (S3), the longitudinal MC $\Delta\sigma_{xx}$ shows hysteresis below 4 K (Fig.\ref{MR} (h)). The hysteretic dips in MC are readily attributed to well-known contributions from domain wall scattering at the coercive field of the ferromagnet.

\begin{figure*}
\includegraphics[width=150mm]{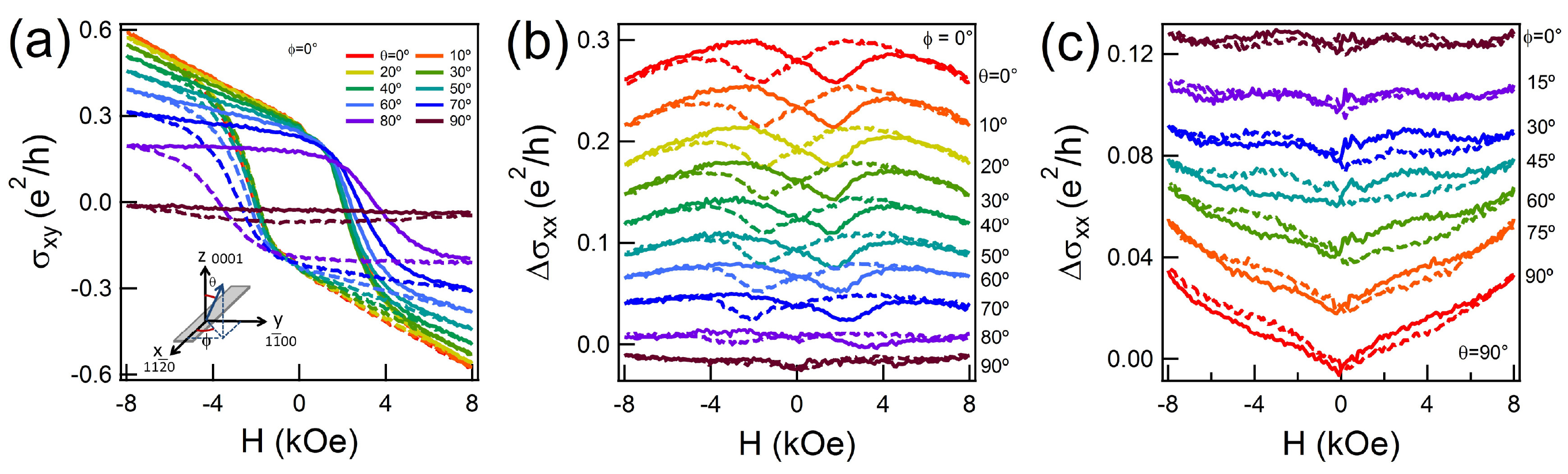} 
\caption{(Color online) Angular dependence of Hall conductivity and longitudinal conductivity of a Mn-doped \BT~thin film with 2 \% Mn concentration (S3) at 0.3 K. A schematic of a channel with cartesian coordinate and the corresponding crystal directions is shown in (a) inset. (a) Transverse conductivity $\sigma_{xy}$ and (b) longitudinal MC $\Delta\sigma_{xx}$ with polar angles of the magnetic field from $\theta=0^{\circ}$ (perpendicular to the plane) to $\theta=90^{\circ}$ (in-plane along channel direction). Solid (dashed) curves are up-sweeps (down-sweeps). (c) Longitudinal MC $\Delta\sigma_{xx}$ with an in-plane magnetic field from $\phi=0^{\circ}$ (along current direction) to $\phi=90^{\circ}$ (perpendicular to the current direction). Offset for clarity.}
\label{angular}
\end{figure*}

We also studied magneto-transport with different directions of the magnetic field using a cryogenic vector magnet system at 300 mK. The angle $\theta$ and $\phi$ are defined as the polar angle along the c-axis of the crystal and the azimuthal angle from a line along the channel direction of Hall bar geometry, respectively (see inset to Fig.\ref{angular}(a)). As the direction of the magnetic field changes from perpendicular-to-the-plane ($\theta=0^{\circ}$) to in-plane along the channel ($\theta=90^{\circ}$ and $\phi=0^{\circ}$), the amplitude of the AHE hysteresis reduces gradually and disappears at $\theta=90^{\circ}$ (Fig.\ref{angular}(a)). The magnetic field dependence of the transverse conductivity $\sigma_{xy}$ is not completely flat at $\theta=90^{\circ}$ due to a slight misalignment (less than $1^{\circ}$) of the Hall bar geometry to the magnetic field coordinate. The angular dependence of the longitudinal conductivity $\Delta\sigma_{xx}$ reveals the evolution of the hysteresis by changing the polar angle $\theta$ with an azimuthal angle $\phi$ fixed to the channel direction as shown in Fig.\ref{angular}(b). As $\theta$ increases from $0^{\circ}$ to $90^{\circ}$, the magnitude of the hysteresis becomes smaller and the peak position broadens, following the coercive field obtained by transverse conductivity $\sigma_{xy}$ at each angle. With a magnetic field parallel to the plane ($\theta=90^{\circ}$), large hysteresis is not shown with azimuth angles (Fig.\ref{angular}(c)). The result of angular dependence of AHE and the longitudinal MC are consistent with SQUID data, which confirms that the easy axis of the Mn-\BT~thin film is perpendicular to the plane along c-axis of the crystal. 

\begin{figure}
\includegraphics[width=80mm]{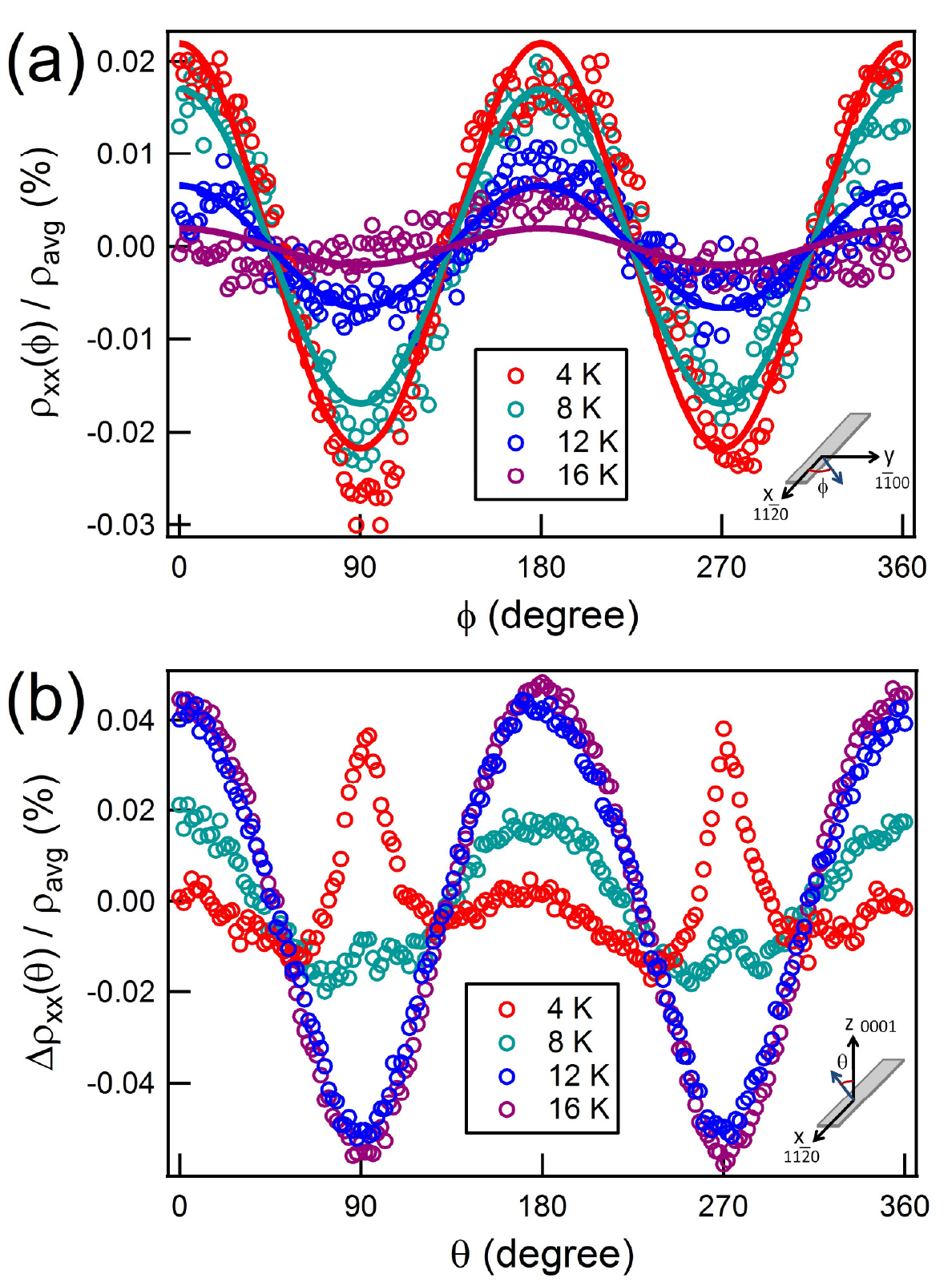} 
\caption{(Color online) Ferromagnetic signatures of AMR in the Mn-\BT~film with 2 \% Mn concentration (S3). Normalized resistivity $\rho_{xx} / \ rho_{avg}$ at a fixed magnetic field (0.5 kOe) (a) with azimuthal angle $\phi$ in the $xy$-plane and (b) with polar angle $\theta$ in the $zx$-plane at different temperatures. In (a), solid lines are fits by Eq. (1).}
\label{AMR}
\end{figure}

The study of angular dependence of magneto-transport reveals the anisotropic magnetoresistance (AMR) in the Mn-\BT~film by showing MC with different directions of the magnetic field sweep with respect to the direction of current (Figs.\ref{angular}(b) and (c)). For further studies of AMR, we carried out magneto-transport measurements with a constant magnetic field of 5 kOe to saturate the magnetization of the Mn-doped \BT~film (S3) and while sweeping through $360^{\circ}$ in the $xy$-plane and the $zx$-plane. In conventional AMR, the magnetoresistivity is expressed as a function of the relative direction of magnetization and current:\cite{OHandley1999} 
\begin{equation}
\rho=\rho_{\|}+[\rho_{\|}-\rho_{\perp}]cos^2\varphi,
\end{equation}
where $\rho_{\|}$ $(\rho_{\perp})$ is the resistivity with a current direction parallel (perpendicular) to the magnetization and $\varphi$ is the angle between magnetization and current. The magnetoresistivity with a rotating in-plane ($xy$-plane) magnetic field fits well with Eq. (1). In Fig.\ref{AMR}(a), the anisotropy weakens as the temperature increases to the ferromagnetic \TC. We note that the magnetoresistivity at 16 K, which is above \TC, is not completely isotropic, which is understood as AMR that is present even in undoped 3D topological insulators.\cite{Wang2012, Zhang2012} AMR of undoped films in the $zx$-plane is known to be bigger than that in the $xy$-plane, which explains the sizable AMR of the Mn-doped \BT~film (S3) in the $zx$-plane at 16 K as shown in Fig.\ref{AMR}(b). Interestingly, the AMR of S3 in the $zx$-plane below \TC~is suppressed as temperature decreases and the enhancement in resistivity is observed as the magnetic field approaches the direction of current ($\theta=90^{\circ}$ and $270^{\circ}$). The AMR results are qualitatively consistent with the results of the angular dependence of the longitudinal MC with magnetic field sweep discussed in this section. The enhancement of resistivity near $\theta=90^{\circ}$ and $270^{\circ}$ in the $zx$-plane is attributed to the domain-wall scattering when the $z$-component of magnetization flips from $+z$ ($-z$) direction to $-z$ ($+z$) direction.

\section{Discussion}
Our study of MBE-grown, Mn-doped \BT~thin films is worth comparing with earlier reports on two distinct types of bulk crystals: Mn-doped \BT~and Mn-doped BiTe.\cite{Hor2010, Bos2006} Up to 2 \% Mn concentration, the crystal structure of the MBE-grown Mn-doped films is almost identical to that of pure Bi$_2$Te$_3$. In the study of Bi$_{2-x}$Mn$_x$Te$_3$ bulk crystals, ferromagnetism was reported over a Mn concentration range of 0.8 $-$ 1.8 atomic percent ($0.04 \leq x \leq 0.09$) with 9 K $\lesssim$ \TC $\lesssim$ 12 K.\cite{Hor2010} Within that range, Mn behaved as an acceptor that substitutes for Bi in the \BT~crystal, and conduction was p-type. This is similar to the values of \TC~in sample S3 except that our sample is n-type, suggesting that carrier-mediated ferromagnetism is not playing a role. At higher Mn concentrations in our study, the crystal structure of our films becomes comparable to that reported for Mn-doped BiTe bulk crystals, which consisted of repeated structures of Bi bilayers between every other \BT~QL, with Mn atoms preferentially localized in the Bi bilayers.\cite{Bos2006} Those bulk crystals showed n-type conduction, as in our samples, and had a \TC~around 10 K for comparable Mn concentrations (6.25 atomic percent). Our key conclusion is that neither crystal nor carrier-type seem to be playing a dominant role in determining the nature of ferromagnetic order in Mn-doped \BT~or BiTe thin films and bulk crystals. 

In conventional diluted magnetic semiconductors, such as (Ga,Mn)As and (In,Mn)As, the ferromagnetic interaction between local moments is mediated via itinerant carriers located in either the valence band or in an impurity band. In a mean field Zener model, \TC~increases with carrier density and with Mn concentration.\cite{MacDonald2005} We do not observe any obvious  dependence of \TC~on carrier density in our n-type Mn-doped \BT~films. IR spectroscopy studies carried out thick Mn-\BT~samples also suggest that charge carriers are not likely mediators of ferromagnetism, showing little change in the IR spectrum upon cooling across \TC.\cite{Chapler2013} It is well known that carrier-density-independent ferromagnetism could possibly arise in an inhomogeneous, phase-separated material that contains either readily observable metallic, ferromagnetic inclusions or nanoscale metallic, ferromagnetic phases.\cite{Dietl_NMat} Our detailed structural studies clearly rule out the former possibility and--to the extent possible with our present microscopy studies--we do not see any obvious signs of the latter. One of the possible interpretations of the carrier-independent ferromagnetism in the Mn-doped \BT~films is that it is of the van Vleck-type where magnetic impurities are coupled by a large magnetic susceptibility of the band electrons.\cite{Yu2010} Ferromagnetism in MBE-grown, Cr-doped (Bi$_x$Sb$_{1-x}$)$_2$Te$_3$ thin films has been attributed to this mechanism.\cite{Chang2013} Another possibility is ferromagnetism mediated by Dirac fermions in the surface states. \cite{Liu2009,Abanin2011} Studies of exfoliated, Mn-doped \BT~crystals where the chemical potential was gated into the bulk band gap saw an increase in the size of the AHE, a signal similar to that observed in Cr-doped samples, and speculated that the ferromagnetic order was mediated by Dirac electrons.\cite{Checkelsky2012} Since the chemical potential in our samples clearly lies in the bulk conduction band, we cannot conclusively address this question. We note, though, that the surface states overlap the conduction band in energy: thus, the presence of Dirac electron-mediated exchange might be expected to result in an enhanced ferromagnetic order at the surface, distinct from ferromagnetism in the bulk. However, our PNR measurements do not show any such signatures, leading us to believe that the van Vleck mechanism is mostly likely responsible for the ferromagnetism observed in our samples.

\section{Conclusions}
In conclusion, we have carried out the synthesis and detailed characterization of magnetically doped TI thin films of Mn-\BT~grown by MBE. Mn doping induces changes in the crystal structure of Bi$_2$Te$_3$, observed by XRD and high-resolution TEM. SQUID magnetometry, as well as measurements of AHE as a function of angle, revealed ferromagnetism below 17 K with a perpendicular-to-plane easy axis. Our results suggest carrier-independent ferromagnetism in Mn-doped \BT~thin films.
Experiments on Mn-doped \BT~films with a wider range of Mn concentrations and with $E_{F}$ in the bulk band gap are required to further confirm the mechanism of the ferromagnetism. Understanding the ferromagnetic properties of Mn-\BT~thin films could possibly provide another pathway to explore the quantum phenomena caused by broken TRS in topological insulator surface states.

{\bf Acknowledgement}

This work was primarily supported by ARO through the MURI program (AR, JT, TCF and NS) and by ONR. NS, JSL and DWR acknowledge additional support through C-SPIN, one of six centers of STARnet, a Semiconductor Research Corporation program, sponsored by MARCO and DARPA. We thank Chao-xing Liu for insightful discussions.


%

\end{document}